\documentclass[conference]{IEEEtran}

\usepackage{cite}

\ifCLASSINFOpdf
 \usepackage[pdftex]{graphicx}

\else

\fi

\begin{document}

\title{Group Maturity and Agility, Are They Connected? \\ -- A Survey Study}

\author{\IEEEauthorblockN{Lucas Gren}
\IEEEauthorblockA{Chalmers and Gothenburg University\\
Gothenburg, Sweden 412--92 and\\
University of S\~ao Paulo\\
S\~ao Paulo, Brazil 05508--090\\
Email: lucas.gren@cse.gu.se}
\and
\IEEEauthorblockN{Richard Torkar}
\IEEEauthorblockA{Chalmers and Gothenburg University\\
Gothenburg, Sweden 412--92 and\\
Blekinge Institute of Technology\\
Karlskrona, Sweden 371--79\\
Email: richard.torkar@cse.gu.se}
\and
\IEEEauthorblockN{Robert Feldt}
\IEEEauthorblockA{
Blekinge Institute of Technology\\
Karlskrona, Sweden 371--79\\
Chalmers and Gothenburg University\\
Gothenburg, Sweden 412--92 and\\
Email: robert.feldt@bth.se}

}

\maketitle

\begin{abstract}
The focus on psychology has increased within software engineering due to the project management innovation ``agile development processes''. The agile methods do not explicitly consider group development aspects; they simply assume what is described in group psychology as mature groups. This study was conducted with 45 employees and their twelve managers (N=57) from two SAP customers in the US that were working with agile methods, and the data were collected via an online survey. The selected Agility measurement was correlated to a Group Development measurement and showed significant convergent validity, i.e., a more mature team is also a more agile team. This means that the agile methods probably would benefit from taking group development into account when its practices are being introduced. 
\end{abstract}

%Agile Processes \sep Measurement \sep Group Psychology \sep Maturity \sep Empirical Study

\IEEEpeerreviewmaketitle

\section{Introduction}
Groups have existed as long as humans have and our ability to form and work in groups is key to our survival and development. However, many people dislike working in groups because group work can be cumbersome and involve conflict, hurt feelings, and inefficiency. The reason why organizations want to organize work in group-form is because when a group is working well, it works extremely well compared to other work methods~\cite{wheelan2012}. This aspect has become more evident in software engineering lately. Before, the traditional approach to software development was that projects usually were considered to be ``plan-driven''. These methods come from the systems engineering and other disciplines, and were established to coordinate large inter-operating components. However, software does not function as hardware and different standards were therefore introduced~\cite{boehmm}.

Group Development research got much attention before the 1960s. After that, the research on groups declined and the focus changed to the individual as the unit of research. Applied studies were still conducted that tried to understand productivity and effectiveness in groups. However, to try to fix groups one must understand how they work. Or as Wheelan and Hochberger~\cite{wheelan} adequately put it: ``before one jumps to fix something, one has to know what is broken''.

When changing to an agile method (e.g. eXtreme Programming, Scrum, Lean etc.), where cooperation and a self-organizing team are central, some aspects of the modern workplace might cause problems. If group members are unable to, e.g., be physically present during meetings, the aspect of human interaction becomes harder to achieve and problems concerning communication, culture, trust, and knowledge management appear~\cite{jala}. Because of the new agile management technique, organizational psychology issues have gotten more attention in software engineering~\cite{bali}. Over the years there have been many models within psychology on how groups behave. There seemed to be a pattern of what happens to all groups. These patterns have been categorized into different stages and labeled differently by many researchers. Bion~\cite{bion}, for example, states that a group always has two states; the work group, and the basic assumption group (consisting of dependency, fight-flight, and pairing stages). Tuckman and Jensen~\cite{tuckman} defined a classic development model with the phases; forming, storming, norming, and performing. These stages perfectly correspond to the theory used in this study and are described in the Group Development Section in this paper. 

The study of agile development is quite a new research field and the connection between agile methods and psychology has just begun. Some studies have been conducted regarding agile methods in connection to group norms~\cite{teh,melnik,moe2010}, culture~\cite{iivari, tolfo2008,whit,tolfo}, personality traits \cite{mcdonald,seger,feldt2010}, and job satisfaction~\cite{melnik2,gren1}, but only one article has been found on agile teams and group psychology~\cite{teh}, in which they conclude that productive group norms give better results. 

Some people might think that focusing on relations in a group is a waste of time. The group must work from the beginning and not spend time on unimportant matters. The problem is that group development cannot be circumvented. The fact is that a mature group performs much better; e.g., they finish projects faster~\cite{wheelan1998}. Students perform better on standardized test if the faculty team is at a mature development stage~\cite{wheelan1999, wheelan2005}, and intensive care staff functioning in a mature team save more lives~\cite{wheelan20032}. Paying attention to group development could help the group to succeed. The agile methods assume that groups are mature since they describe agile collaboration in the same way as a characteristic mature group. The agile methods cannot skip the aspect of group development, since all groups will have to deal with its effect, i.e., much more of what happens in agile teams would be understood better if group development was taken into account. However, in order to justify such a merge, it must be shown that agility and group development have connections. 

The main contribution of this study is to investigate if there is a correlation between two independent measurements of agility and group development. We already want to highlight the importance of separating between correlation and causality here. We can not say if agile processes drive teams to mature or if mature teams are needed to adopt agility by doing a correlation study. What we can say is that the concept are connected and the one coexists with the other.

We have not found many thoroughly validated agile measurements, and we selected one of the well-cited tools to measure ``agility'', based on the overview presented by Leppanen~\cite{lepp}. One issue is the definition of agility since we need to know what to measure. The reason why we chose Sidky's~\cite{sidkyphd} is that it provides a set of items in survey form that aims to measure the behavior connected to agile processes. The tool was validated with expert practitioners regarding its content, but has not been statistically validated (few of the agility measurements are~\cite{lepp}). If we found a correlation we, at least, can conclude that there are indications that the concept of agility is connected to the group maturity level, since practitioners have confirmed that the tools measure what they consider ``agility'' to consist of. It is, of course, evident that agile teams also go through the same development as other teams. However, if there is a correlation and an agile team has many similarities to a mature group, the implementation of agile methods could be adapted to the group's maturity level, which could increase the understanding of agility and why they are easier to implement in some contexts. This will also partly help define agility since group maturity actually is one of the dimensions of agility.

%\subsection{Hypothesis}

\paragraph{Hypothesis} The agility measurement has significant convergent validity with regards to the group development measurement. 

Section~\ref{sec:related_work} will outline group development research and present the agility measurement used, Section~\ref{sec:methodology} will present the survey and the statistical investigation conducted. Section~\ref{sec:discussion} will analyze and discuss these results, and, finally, Section~\ref{sec:conclusions} will present conclusions and suggest future work.

\section{Related Work}\label{sec:related_work}

\subsection{Groups and Teams}
Keyton~\cite{grupp} defines a group as: ``three or more members that interact with each other to perform a number of tasks and achieve a set of common goals'', which means that large groups are in fact a set of smaller subgroups and should be considered separately. If a group consists of more than eight individuals it is less productive than a smaller group~\cite{wheelan2009}. A ``work group'' consists of members that want to create a shared view of goals and develop a structure to achieve the goals. A ``team'' is a work group that has shared goals and effective methods to achieve them, according to~\cite{wheelan}. This implicates that many work groups in organizations are not teams. Only 17\% of all groups were considered teams according to one study~\cite{wheelan}.

A modern aspect of teams is team building. Sadly, team building has been mixed up with getting to know people on a personal level. If a set of individuals are hired to work together to solve a certain problem, they will only benefit from getting training in the process that leads to the solution. If they spend their spare time and become life-long friends that may be good but not always necessary to get high performance teams at work~\cite{wheelan2012}. Team building activities are of utter importance to team efficiency if they are connected to the group goal and the process of getting there~\cite{klein}.

\subsection{Wheelan's Integrated Model of Group Development}\label{sub:integratedgroup}
\begin{figure*}
\centerline{\includegraphics[width=130mm]{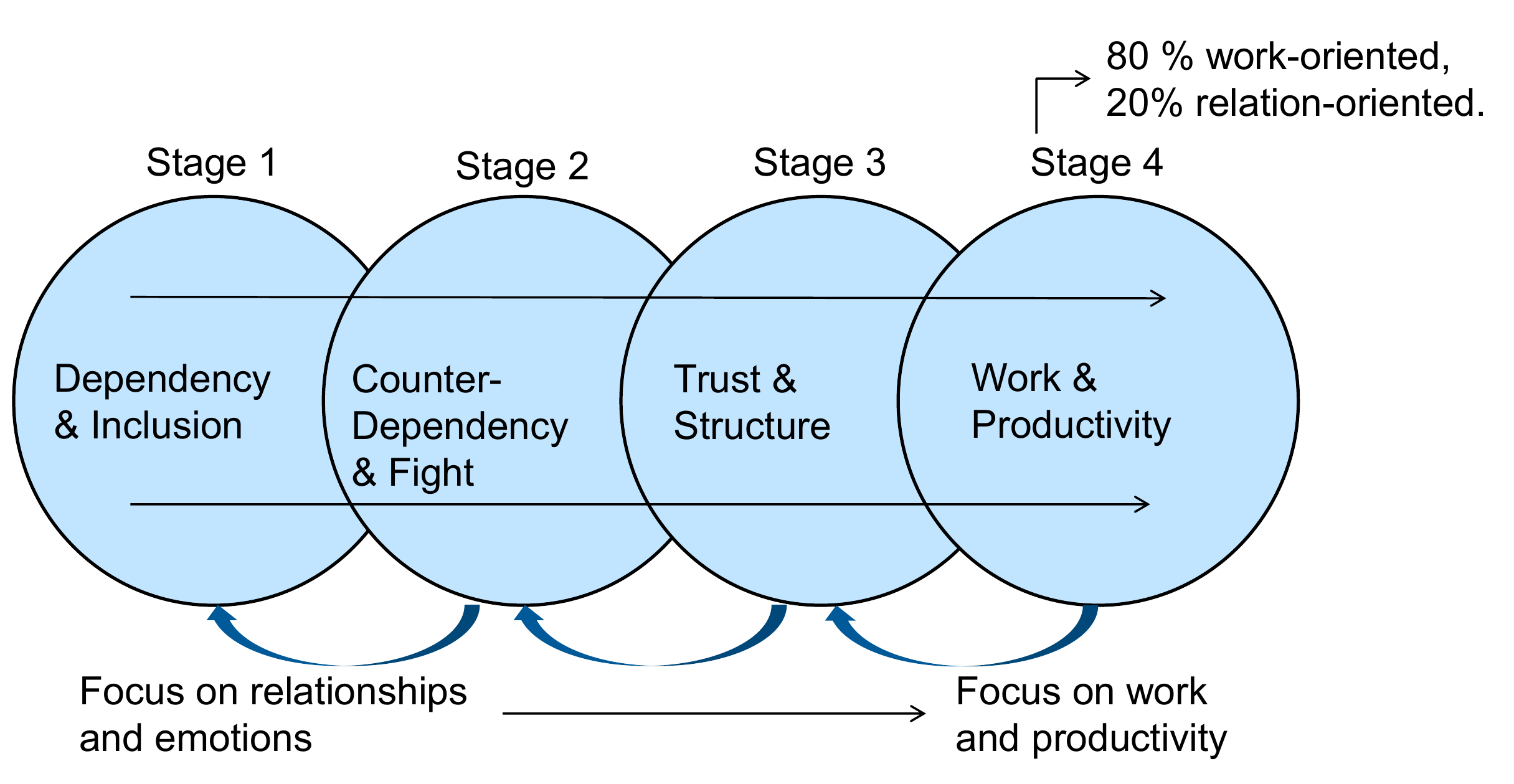}}
\caption{The Group Development Stages~\cite{wheelan2012}}
\label{fig:groupstages}
\end{figure*}

Many group development theories describe a dynamic view of a group. Older theory like the one presented by Bion~\cite{bion} as well as newer group dynamic theory all evolve around a set of stages that groups go through~\cite{wheelan1993}. The theory used in this study presented by Wheelan and Hochberger~\cite{wheelan} is actually an integrated model of group development and is also branded as such (the model is called The Integrated Model of Group Development, or IMGD). Wheelan and Hochberger~\cite{wheelan} later connected a survey to this model, called the Group Development Questionnaire (or GDQ). This tool measures the maturity level of a group in 4 different stages (see Figure~\ref{fig:groupstages}). These 4 stages which will be presented in more detail next and the Group Development Questionnaire will be explained in detail afterward.

\paragraph{Stage 1: Dependency and Inclusion}
The first stage is categorized by three main areas; concerns about safety and inclusion, member dependency on the designated leader, and a wish for order and structure. The group is supposed to become organized, capable of efficient work, and achieve goals, so the first state must have a purpose in getting there. The first step is to create a sense of belonging and lay the foundation of predictable patterns of interaction. After the stage one the group members should feel that they belong to that group of people and feel safe enough to contribute with their ideas and suggestions that they believe with help the group reach its goal. If this does not happen a group will stagnate, which happens when people stop attending meetings or do group-related work between them~\cite{wheelan}. 

\paragraph{Stage 2: Counter-Dependency and Fight}
The second stage of a group's development is a conflict phase where fight is a must to create clear roles to be able to work together in a constructive way. The members have to go through this in order to be able to trust each other and the leader. When the group safely navigated through the first stage they have gained a sense of loyalty. As people feel more safety they will dare to speak up and express opinions that might not be shared by all members. In the second stage some hard work needs to be done. As was described about culture in earlier the shared perceptions of values, norms, and goals need to be set for the group. This set of rules need to be negotiated and put in place for the group to work effectively. In the first stage, some primary discussions about how people should interact and organize probably surfaced, however on a shallow level. More work is needed though in order to agree on these aspects and every member needs to participate for this to occur~\cite{wheelan}.

\paragraph{Stage 3: Trust and Structure}
The third stage is structure-developing phase where the roles are based on competence instead of striving for power or safety. Communication will be more open and task-oriented. The third stage of group development is characterized by more mature negotiations about roles, organization, and processes. The second category dealt with is the solidification of positive relationships. When the groups negotiate roles, organization, and procedures there will be an evident clarification and consensus about goals. The roles and tasks will also be adjusted to make goal achievement likelier. The leader's role will be more consultative and less directive at this stage, and the communication structure is more flexible. The content of the communication is also more task-oriented instead of relation-oriented. Here, a greater tolerance of subgroups, cliques, and coalition is evident, and there is also an increased division of labor. Conflicts continue to occur but the conflict management is more effective. When the group builds the positive relationships there is an increase in the group cohesion and trust. The group members will also cooperate more and their work satisfaction will increase. Individual commitment to group goals and task will be high, and voluntary conformity with the group norms upsurges. Deviation from the group will be accepted if helpful for the group~\cite{wheelan2012}. 

\paragraph{Stage 4: Work and Productivity}
The fourth and final stage (excluding the termination phase) is when the group wants to get the task done well at the same time as the group cohesion is maintained over a long period of time. However, a group that only works is not the most effective one. Groups that work 80\% and puts 20\% into work-related relations (dealing with conflict and intra-personal issues) are the most effective ones. The group also focuses on decision-making and encourages task-related conflicts. This is a time of intense productivity and effectiveness. It is at this stage the group becomes a team. Of course, work is done at all stages but here there is a vast increase in focus on task accomplishment. Many people have never been a part of a stage four team. Some people that get to experience this say they never got to it before. There is usually an excitement in enjoying work so much and an ease in getting work done. Members of a team are usually thrilled and want the work to continue that way for as long as possible. This is possible, but it is also likely that backsliding occurs if the team is not careful. Getting to stage four is not easy and many groups never get there. The team needs to know how to stay in stage four~\cite{wheelan2012}.

\subsection{The Group Development Questionnaire (GDQ)}
In our opinion, the largest contribution by Wheelan~\cite{wheelan2012} was to connect a questionnaire to group development. In doing so it has become possible to diagnose and pinpoint the current maturity level of a group. Their survey has a total of 60 items and provides a powerful tool for research and interventions in teams. The Group Development Questionnaire is divided into four different parts. Each of them measure how much energy is put into each stage of group development. The fourth part (GDQ4) measures work and productivity and has been shown to correlate with a diversity of effectiveness measures in different sectors, e.g., groups that have high scores on GDQ4 finish projects faster~\cite{wheelan1998}, students perform better on standardized test if the faculty team is has high scores on GDQ4~\cite{wheelan1999, wheelan2005}, and intensive care staff saves more lives~\cite{wheelan20032}.

\subsection{Agility Measurement}
Sidky's Agility Measurement Index (or SAMI) is a tool developed by Sidky~\cite{sidkyphd}. In Sidky's \cite{sidkyphd} framework, the researcher\slash assessor conduct qualitative interviews and set up numbers 1-5 in a questionnaire depending on the answers. So, Sidky's~\cite{sidkyphd} framework is a mix between qualitative and quantitative research, where the researcher transforms qualitative answers into quantitative numbers (in this study, though, we let the group members fill out the survey and allows a statistical evaluation of the result). Sidky~\cite{sidkyphd} validated the content in the tool by letting practitioners evaluate the items and their connection to what they think agility is. 

Sidky divided agile practices into different levels but in order to keep the items to a feasible number we only used Level 1 in this study (see Sidky~\cite{sidkyphd} for more details). He also divides these according to the agile principles: ``Embrace change to deliver customer value'', ``Plan and deliver software frequently'', ``Human-centric'', ``Technical excellence'', and ``Customer collaboration'' of the agile manifesto~\cite{fowler2001}. 

To clarify, this means that the practices ``Reflect and tune process'', ``Collaborative planning'', ``Collaborative teams'', ``Empowered and motivated teams'', ``Working standards\slash procedures'', ``Knowledge sharing tools'', ``Task volunteering'', and ``Customer commitment'' were measured in this study. The actual items (or indicators) are left out but can be found in Sidky~\cite{sidkyphd}. The measurement tool consists of one survey for managers and one survey for developers. To assess an agile practice the analysis method proposed uses answers from both surveys. Some of the items are also used to assess more than one practice.

\section{Method}\label{sec:methodology}
This section presents the method used to assess agility, group maturity, and correlate these measurements.

\subsection{Participants}\label{sub:participants_all}
The survey sample consisted of 45 employees and their twelve managers ($N=57$) from two large multinational US-based companies with 16,000 and 26,000 employees and with revenues of US\$ 4.4 billion and US\$ 13.3 billion respectively. Both stated that they are using agile methods in the projects that participated. One of the companies is in the Retail Business and the other is in the Consumer Packaged Goods (CPG) industry. However, the groups participating in the research were only IT projects. This study was conducted together with SAP America Inc.\ and they mediated the contacts, however, only two of the participating teams were Agile SAP implementation projects. 

%One of the companies introduced agile methods a while back in order to have more ownership and buy-in from staff along with increased job satisfaction. 

\subsection{Surveys}\label{sec:methodologysurvey}
The surveys used in this study were Sidky's~\cite{sidkyphd} manager and developer surveys from agile level 1 presented earlier with part 4 (GDQ4 Work and Productivity) of the Group Development Questionnaire~\cite{wheelan} added in the beginning of both surveys. The Agile Measurement surveys (separate manager and developer parts and both with the GDQ4 in the beginning) were put together in online surveys containing 41 and 44 items respectively for the team members and managers to answer on a Likert scale from 1 to 5 (where 1 = low agreement to the statement and 5 = high agreement). Below are three sample question from the GDQ4 and three from the agility measurement.

\begin{itemize}
\item GDQ4: The group gets, gives, and uses feedback about its effectiveness and productivity.
\item GDQ4: The group acts on its decisions.
\item GDQ4: This group encourages high performance and quality work.
\item Agility: Process change in the middle of the project should not be considered a disruption since the process change is worth the benefit it will bring.
\item Agility: There should be a mechanism for persistent knowledge sharing between team members.
\item Agility: When you run into technical problems, you usually ask your team members about the solution. 
\end{itemize}

In order to say which group stage a group is in, the whole 60-item GDQ survey must be used. However, it is possible to only measure the degree of effective group work by only using scale four of GDQ\@. This reduces the number of items to 15, which makes it less time consuming. This is the scale that was correlated to other effectiveness measurements in health care, schools etc.\ (see Introduction Section).

\subsection{Procedure}
The surveys were sent out to the employees via email by their manager. The survey was created as an online survey and the link to it was shared in the email. The surveys were sent to 92 employees and 57 replied, i.e., a response rate of 61\%. This response rate is above average (55.6\%) within social science research~\cite{responserate}. One reminder was sent via email by one of the managers (from one of the organizations). Filling out the survey took approximately 10 minutes and all the questions were compulsory. They were sent to the employees by the manager who stated how many months each participating group had been working together.

\subsection{Data Analysis}
A first step to see if there is a connection between agility and group development would be to do a simple correlation analysis of the same individual's mean value results in the two measurements. Assuming that the agile tool really measures some aspects of agility, this would show that they are connected. 

\begin{figure}
\centerline{\includegraphics[width=90mm]{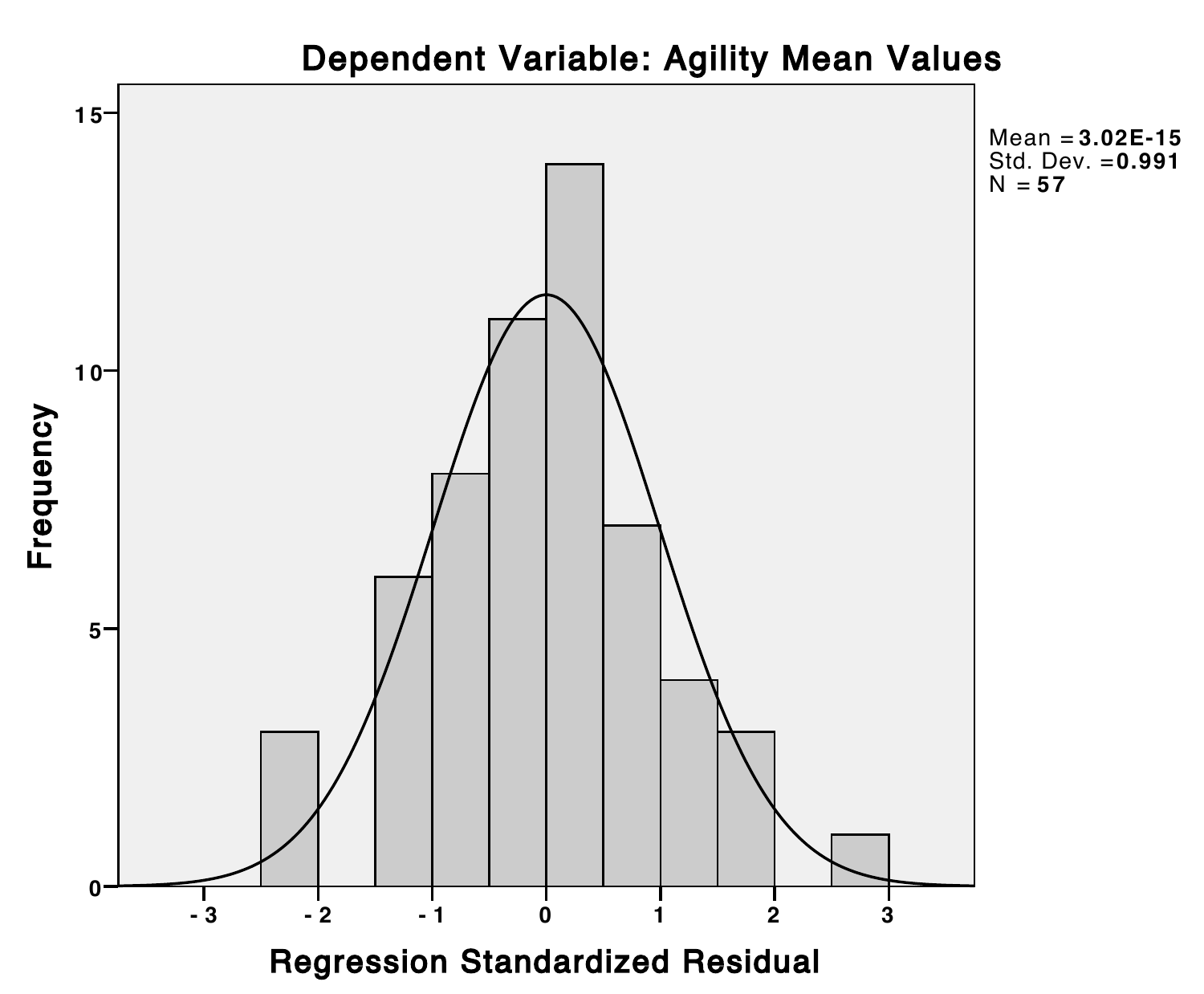}}
\caption{Histogram to Check the Normality Assumption for our Linear Model.}
\label{fig:hist}
\end{figure}

In order to evaluate if the data was normally distributed we plotted the frequency of the Regression Standardizes Residuals (see Figure~\ref{fig:hist}). We also conducted a Shapiro-Wilk Test for each value of the GDQ4. As Figure~\ref{fig:hist} shows there were no surprise that none of them were significant, i.e.\ our normality assumption is valid for our linear regression model. 

\section{Results}\label{sec:results}
\subsection{Results from the Surveys} 
In order to calculate the intervals to compare to nominal scores, the items belonging to each category were calculated according to the procedure described in Sidky~\cite{sidkyphd}, with one small alteration to the tool; the result of the item ``Other peoples' titles and positions intimidate people in the organization''. This variable was inverted because the aspect of intimidation of titles must be seen as a negative thing within the agile approach. Sidky~\cite{sidkyphd} also writes that this item is used to determine: ``Whether or not people are intimidated\slash afraid to give honest feedback and participation in the presence of their managers''. Sidky also confirmed this himself in email cadence.

A problem with using the SAMI to measure agility is the terms ``Manager'' and ``Scrum Master\slash Agile Coach''. Two of the respondents gave the following feedback: ``We have a PM and an Agile coach. I consider their agile skills to be far apart which leads to some ambiguity when answering questions around `manager'.'', and ``Some of the questions on my manager are irrelevant or could be misinterpreted. My manager isn't part of the IT organization.''. This ambiguity probably affects the responses since some of the agile teams have both a Manager and a Scrum Master. A purely agile team is not supposed to have a manager and to have both managers and agile leaders is not part of any agile framework. As mentioned before, the Scrum Master manages the project but the team is supposed to be organizing itself. This will make it harder for us to find significant results since the responses will contain more variance. However, a mixture of methods is the reality in many companies today~\cite{west}.

The amount of time the groups had been working together ranged from 1--24 months and with a mean value of 7 months.

\subsection{Results from Statistical Analyzes}\label{sec:resultsall}

The result of this study is under the assumption that Sidky's~\cite{sidkyphd} SAMI items reflect some aspects of agility. 

\begin{figure*}
\centerline{\includegraphics[scale=0.7]{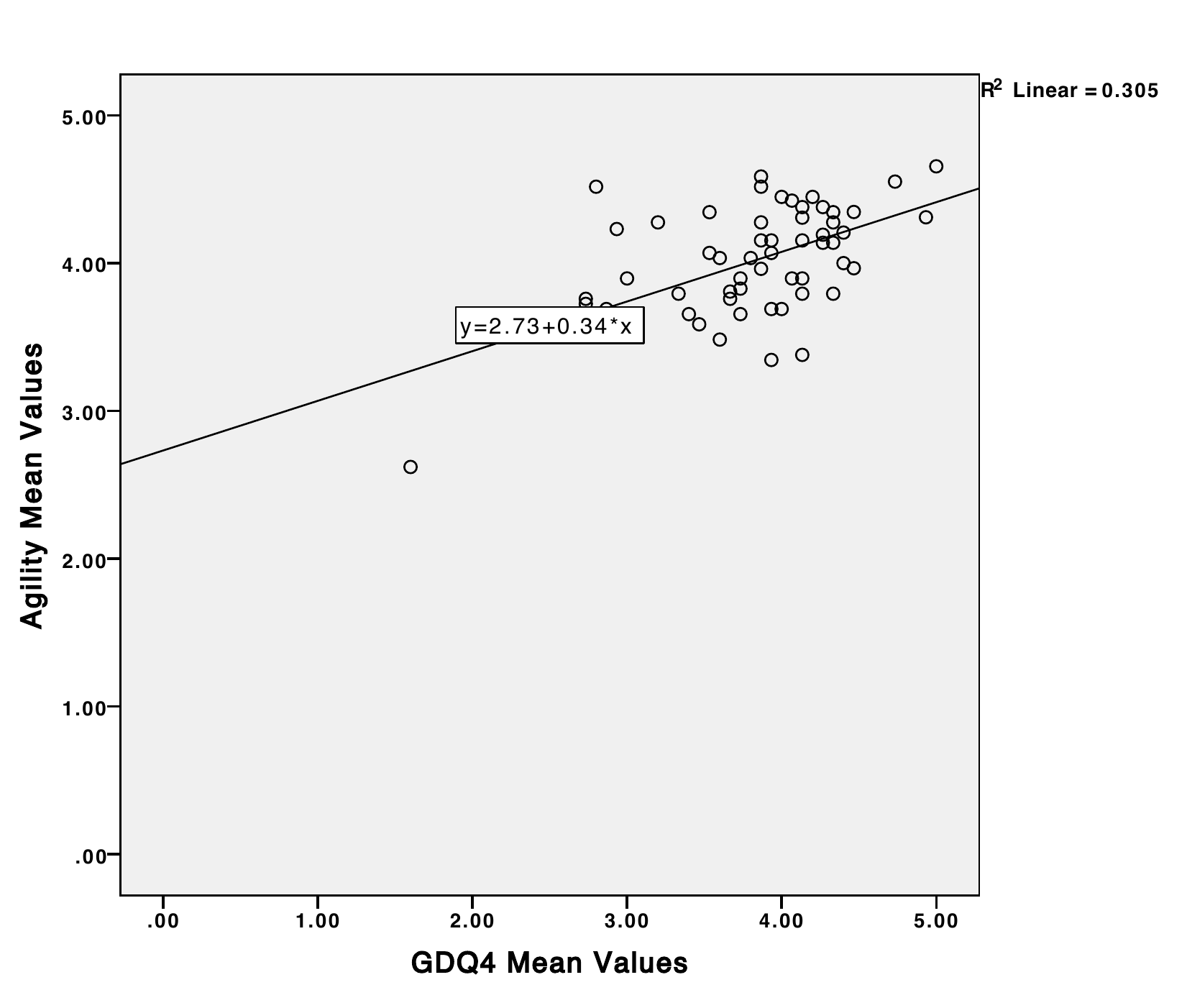}}
\caption{Scatterplot with ``Agility'' Mean Values as Dependent Variable and the Group Development Mean Values as Independent Variable}
\label{fig:scatterdot}
\end{figure*}

Figure~\ref{fig:scatterdot} shows the scatter-plot for all the agile practices' mean values as dependent variable and the GDQ4 mean values as independent, for all 57 individuals. Each dot represents one individual and its score (mean value) on both the agility ($y$-axis) and group development ($x$-axis) items. The $R^2$ (i.e.\ explained variance by the model) is 0.305 which means that 31\% of the variance in the agile measurement can be explained by the GDQ4 measurements. 

The data points are evenly spread out around the regression line, which means that the residuals can be  assumed to be reasonably normally distributed. This is an important prerequisite for correlating variables with these types of methods. Table~\ref{fig:regcoll} shows the simple linear regression analysis behind the scatter-dot plot. In a simple linear regression analysis the standardized beta is the same as the Pearson correlation coefficient, i.e., $r = .553$, $p = .000$, which is a strong correlation.

\begin{table*}
\caption{Simple Linear Regression with the ``Agility'' Mean Values as Dependent Variable and the the Group Development Mean Values as Independent Variable}
\label{fig:regcoll}
\centerline{\includegraphics[scale=0.7]{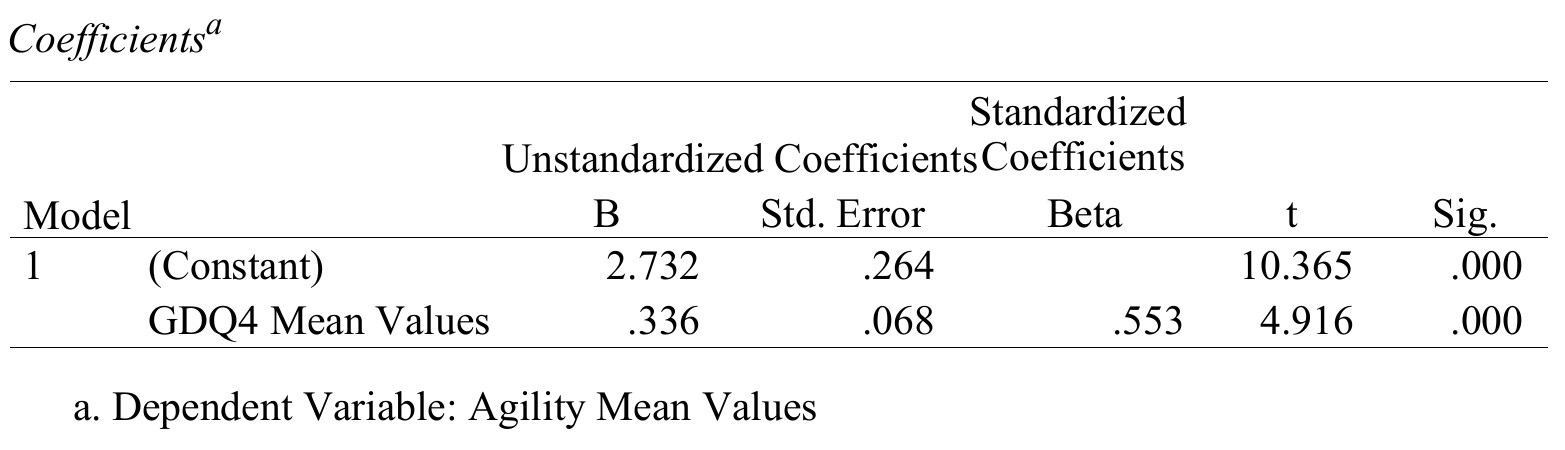}}
\end{table*}

This shows that, in this data set, the GDQ4 measurement can significantly explain variance in the agility measurement and one unit increase in this variable is a $.336$ increase in agility, i.e.\ if a one takes one step from 3 to 4 on variable GDQ4 the variable ``agility'' would increase by $.336$. 

We also did a $t$-test to see if there were any differences between how managers and developers assessed the agility level. We found no such difference ($t=1.039, p=.303)$.

\section{Discussion}\label{sec:discussion}
The results of this study show that the measurements for group development and agility we used are correlated. This is a first step to evaluate the usefulness of further studies and understanding within this new research sub-area. In the long run, this could have implications for how to approach agile methods and suggest that parts of the agile approach possibly are aspects of group development. This gives a positive answer to the research hypothesis whether the Agility measurement has significant convergent validity with the Group Development measurement.

The fact that they are connected shows that agile methods are probably harder to implement in new teams than traditional methods. Or, since the group needs to be mature to be agile, the methods might also enable the group to develop. Since there is an important difference between correlation and causation we need to reflect on all possible reasons for our found correlation. First, agility causes groups to mature, second, mature groups are needed to have agility, third, other unknown variables mediate both high agility and group maturity.

The first possible causation (that agility pushes groups to mature) is not supported by earlier research according to us. According to, for example,~\cite{doingtobeing, williams} teams can implement agile practices without being agile. Even if agility is a concept that is hard to define, the agile practices can be seen as enablers for group maturity, but are not enough on their own. Of course, if one defines agility as a culture that is responsive to change the true agile team is a stage four high-performing team as well.  

The second possible causation (that team maturity is a prerequisite for agility) also comes back to the definition. If we mean agile practices here, teams do not have to be a stage four team to be agile. However, we believe that a culture that is responsive to change in absolute terms (and not in comparison to an earlier plan-driven\slash waterfall management approach) do need to build a mature team (with productive groups norms, good conflict resolution and decision-making techniques etc.) in order to being able to work in such a flexible way.

The third option (that other variables affect both our measurements) is very likely in a complex adaptive systems such as a team in an organization. What is interesting with the GDQ measurement is that it measures quite abstract group development patterns the reflect the team's internal perception of cooperation as well as external factors that affect the team. Of course there are a large set of other variables that influence both our measurements and we can not either be sure that our measurements reflect the underlying construct we aim to investigate. Maybe the group development in an agile context needs to be measured differently and be intertwined in a new way for the process to be understood from a causality point of view. This is research we are most interested in conducting in the future. 

In the following part of this discussion we will give recommendations under the assumption that team maturity somehow is a prerequisite for agility.

The impact of group norms are known, but not given enough attention in software engineering research to be able to guide both practitioners and researcher, according to us (except for one article published 2012 \cite{teh}, presented in the Introduction Section). The fact that agility and group development are correlated actually strengthens the agile approach since these measurements are also correlated to other effectiveness measurements in other fields~\cite{wheelan1998, wheelan1999, wheelan2005, wheelan20032}. In other words, this study helps define part of agility as being aspects of group development. Maybe the methods can push teams into becoming high performing in a way that traditional plan-driven project management techniques can not. If the agility measurement is positively correlated to the group maturity measurement it would be expected that waterfall methods are the opposite. The results presented in this study should be seen as an indication that project management ideas have much to learn from social and group psychology when it comes to helping teams to mature. For example, by integrating the dynamic and developmental view of ``goals'', ``roles'', ``interdependence'', ``leadership'', ``communication and feedback'', ``discussion, decision-making, and planning'', ``implementation and evaluation'', ``norms and individual differences'', ``structure'', and ``cooperation and conflict management'', the processes of building an agile team will be clearer. It is also only lately this type of measurements are possible with high validity~\cite{wheelan} and, thus, a lot of work conducted in teams should take the group's maturity level into account. Although all types of projects might not need an agile way of working to meet its requirements, nor have we looked into the effectiveness of the methods. One way to do this could be to monitor velocity on a diversity of teams in connection to their group maturity. 

Since the Group Development Questionnaire has a work psychology intervention connected to it, it would be very interesting to see if groups adopt agile development processes better and faster in groups that get coaching only with regards to their group development. Such an intervention has no connection to Software Engineering today.

The groups participating in this study had been working together from one to 24 months and with an average of seven months. In group development theory a group needs, as a mean value, 6 months to become high performing~\cite{wheelan2009}. This study therefore shows that groups that have met less than 6 months are more unlikely to be high performing and therefore less agile in their processes.

\subsection{Validity Threats}
In this study we have shown that the two different measurements of agility and group development are correlated. Mature groups could function better no matter what the process is and we did not have a control group with traditional waterfall projects. Since agility is positively correlated to group development, traditional teams should not be, however, this is an assumption and should be researched empirically. The agility measurement is also not thoroughly validated, which could mean that it does not even measure ``agility''. This is a problem since agility is very much an ambiguous concept, but we still argue that parts of what is meant by agility is present in the SAMI items. We also neglect the aspect of if the effectiveness of the agile practices used. In the future such a measurement (like velocity) would add the much needed aspect of if these methods are relevant and in which contexts. 

We do believe that an agile group is a mature group, however, we want to be careful generalizing our results, since we have a small sample without much background information about the teams. This study needs to be replicated to empirically provide more evidence before we can draw conclusions about the wider population of agile teams.

\section{Conclusions and Future Work}\label{sec:conclusions}
This paper set out to see if the agility measurement had significant convergent validity with a measurement of group maturity. Through a correlation analysis, we found such a correlation using a simple linear regression analysis. These findings are important contributions to agile software development because an agile team seems to also be a mature team as defined by social psychology. We believe, therefore, that aspects and experiences from group psychology should be integrated into the transition to agile software development. 

For future research, we suggest these preliminary results are further investigated with a larger sample. Also, a deeper analysis of what aspects in particular that correlate more strongly or weakly would be interesting to the community. It would also be useful to more rigorously define how the agile methods can be adapted to the group's development stage before they are implemented. It would also be interesting to see how agile practices are adopted over time in relation to group maturity. More studies should also be conducted that include a larger sample size.

\section*{Acknowledgements}
This study was conducted jointly with SAP AG (http://www.sap.com), and we would especially like to thank Jan Musil at SAP America Inc. We would also like to thank the SAP customers who were willing to share information.

\bibliographystyle{IEEEtran}

\bibliography{references}

\begin{thebibliography}{10}
\providecommand{\url}[1]{#1}
\csname url@samestyle\endcsname
\providecommand{\newblock}{\relax}
\providecommand{\bibinfo}[2]{#2}
\providecommand{\BIBentrySTDinterwordspacing}{\spaceskip=0pt\relax}
\providecommand{\BIBentryALTinterwordstretchfactor}{4}
\providecommand{\BIBentryALTinterwordspacing}{\spaceskip=\fontdimen2\font plus
\BIBentryALTinterwordstretchfactor\fontdimen3\font minus
  \fontdimen4\font\relax}
\providecommand{\BIBforeignlanguage}[2]{{%
\expandafter\ifx\csname l@#1\endcsname\relax
\typeout{** WARNING: IEEEtran.bst: No hyphenation pattern has been}%
\typeout{** loaded for the language `#1'. Using the pattern for}%
\typeout{** the default language instead.}%
\else
\language=\csname l@#1\endcsname
\fi
#2}}
\providecommand{\BIBdecl}{\relax}
\BIBdecl

\bibitem{wheelan2012}
S.~Wheelan, \emph{Creating effective teams: a guide for members and leaders},
  4th~ed.\hskip 1em plus 0.5em minus 0.4em\relax Thousand Oaks: SAGE, 2013.

\bibitem{boehmm}
B.~Boehm and R.~Turner, ``Management challenges to implementing agile processes
  in traditional development organizations,'' \emph{Software, IEEE}, vol.~22,
  no.~5, pp. 30--39, 2005.

\bibitem{wheelan}
S.~Wheelan and J.~Hochberger, ``Validation studies of the group development
  questionnaire,'' \emph{Small Group Research}, vol.~27, no.~1, pp. 143--170,
  1996.

\bibitem{jala}
S.~Jalali and C.~Wohlin, ``Agile practices in global software engineering--a
  systematic map,'' in \emph{International Conference on Global Software
  Engineering (ICGSE)}.\hskip 1em plus 0.5em minus 0.4em\relax IEEE, 2010.

\bibitem{bali}
V.~Balijepally, R.~Mahapatra, and S.~Nerur, ``Assessing personality profiles of
  software developers in agile development teams,'' \emph{Communications of the
  Association for Information Systems}, vol.~18, no.~1, p.~4, 2006.

\bibitem{bion}
W.~Bion, \emph{Experiences in groups: and other papers}.\hskip 1em plus 0.5em
  minus 0.4em\relax London: Routledge, 1992.

\bibitem{tuckman}
B.~Tuckman and M.~Jensen, ``Stages of small-group development revisited,''
  \emph{Group \& Organization Management}, vol.~2, no.~4, pp. 419--427, 1977.

\bibitem{teh}
A.~Teh, E.~Baniassad, D.~Van~Rooy, and C.~Boughton, ``Social psychology and
  software teams: Establishing task-effective group norms,'' \emph{Software,
  IEEE}, vol.~29, no.~4, pp. 53--58, 2012.

\bibitem{melnik}
G.~Melnik and F.~Maurer, ``Direct verbal communication as a catalyst of agile
  knowledge sharing,'' in \emph{Agile Development Conference, 2004}.\hskip 1em
  plus 0.5em minus 0.4em\relax IEEE, 2004, pp. 21--31.

\bibitem{moe2010}
N.~Moe, T.~Dings{\o}yr, and T.~Dyb{\aa}, ``A teamwork model for understanding
  an agile team: A case study of a scrum project,'' \emph{Information and
  Software Technology}, vol.~52, no.~5, pp. 480--491, 2010.

\bibitem{iivari}
J.~Iivari and N.~Iivari, ``The relationship between organizational culture and
  the deployment of agile methods,'' \emph{Information and Software
  Technology}, vol.~53, no.~5, pp. 509--520, 2011.

\bibitem{tolfo2008}
C.~Tolfo and R.~Wazlawick, ``The influence of organizational culture on the
  adoption of extreme programming,'' \emph{Journal of systems and software},
  vol.~81, no.~11, pp. 1955--1967, 2008.

\bibitem{whit}
E.~Whitworth and R.~Biddle, ``The social nature of agile teams,'' in
  \emph{Agile Conference (AGILE), 2007}.\hskip 1em plus 0.5em minus 0.4em\relax
  IEEE, 2007, pp. 26--36.

\bibitem{tolfo}
C.~Tolfo, R.~Wazlawick, M.~Ferreira, and F.~Forcellini, ``Agile methods and
  organizational culture: Reflections about cultural levels,'' \emph{Journal of
  Software Maintenance and Evolution: Research and Practice}, vol.~23, no.~6,
  pp. 423--441, 2011.

\bibitem{mcdonald}
S.~McDonald and H.~Edwards, ``Who should test whom?'' \emph{Communications of
  the ACM}, vol.~50, no.~1, pp. 66--71, 2007.

\bibitem{seger}
T.~Seger, O.~Hazzan, and R.~Bar-Nahor, ``Agile orientation and psychological
  needs, self-efficacy, and perceived support: a two job-level comparison,'' in
  \emph{Agile, 2008. AGILE'08. Conference}.\hskip 1em plus 0.5em minus
  0.4em\relax IEEE, 2008, pp. 3--14.

\bibitem{feldt2010}
R.~Feldt, L.~Angelis, R.~Torkar, and M.~Samuelsson, ``Links between the
  personalities, views and attitudes of software engineers,'' \emph{Information
  and Software Technology}, vol.~52, no.~6, pp. 611--624, 2010.

\bibitem{melnik2}
G.~Melnik and F.~Maurer, ``Comparative analysis of job satisfaction in agile
  and non-agile software development teams,'' in \emph{Extreme Programming and
  Agile Processes in Software Engineering}.\hskip 1em plus 0.5em minus
  0.4em\relax Springer, 2006, pp. 32--42.

\bibitem{gren1}
L.~Gren, R.~Torkar, and R.~Feldt, ``Work motivational challenges regarding the
  interface between agile teams and a non-agile surrounding organization: A
  case study,'' in \emph{Agile Conference (AGILE), 2014}, July 28—-August 1
  2014, pp. 11–--15.

\bibitem{wheelan1998}
S.~Wheelan, D.~Murphy, E.~Tsumura, and S.~F. Kline, ``Member perceptions of
  internal group dynamics and productivity,'' \emph{Small Group Research},
  vol.~29, no.~3, pp. 371--393, 1998.

\bibitem{wheelan1999}
S.~Wheelan and F.~Tilin, ``The relationship between faculty group development
  and school productivity,'' \emph{Small group research}, vol.~30, no.~1, pp.
  59--81, 1999.

\bibitem{wheelan2005}
S.~Wheelan and J.~Kesselring, ``Link between faculty group: Development and
  elementary student performance on standardized tests,'' \emph{The journal of
  educational research}, vol.~98, no.~6, pp. 323--330, 2005.

\bibitem{wheelan20032}
S.~Wheelan, C.~N. Burchill, and F.~Tilin, ``The link between teamwork and
  patients' outcomes in intensive care units,'' \emph{American Journal of
  Critical Care}, vol.~12, no.~6, pp. 527--534, 2003.

\bibitem{lepp}
M.~Lepp{\"a}nen, ``A comparative analysis of agile maturity models,'' in
  \emph{Information Systems Development}.\hskip 1em plus 0.5em minus
  0.4em\relax Springer, 2013, pp. 329--343.

\bibitem{sidkyphd}
A.~Sidky, ``A structured approach to adopting agile practices: The agile
  adoption framework,'' Ph.D. dissertation, Virginia Polytechnic Institute and
  State University, 2007.

\bibitem{grupp}
J.~Keyton, \emph{Communicating in groups: building relationships for group
  effectiveness}.\hskip 1em plus 0.5em minus 0.4em\relax New York: McGraw-Hill,
  2002.

\bibitem{wheelan2009}
S.~Wheelan, ``Group size, group development, and group productivity,''
  \emph{Small Group Research}, vol.~40, no.~2, pp. 247--262, 2009.

\bibitem{klein}
C.~Klein, D.~Diaz~Granados, E.~Salas, H.~Le, S.~Burke, R.~Lyons, and
  G.~Goodwin, ``Does team building work?'' \emph{Small Group Research},
  vol.~40, no.~2, pp. 181--222, 2009.

\bibitem{wheelan1993}
S.~Wheelan and R.~Mckeage, ``Developmental patterns in small and large
  groups,'' \emph{Small Group Research}, vol.~24, no.~1, pp. 60--83, 1993.

\bibitem{fowler2001}
M.~Fowler and J.~Highsmith, ``{The Agile Manifesto},'' In Software Development,
  Issue on Agile Methodologies, last accessed on December 29th, 2006, Aug.
  2001.

\bibitem{responserate}
Y.~Baruch, ``Response rate in academic studies-a comparative analysis,''
  \emph{Human relations}, vol.~52, no.~4, pp. 421--438, 1999.

\bibitem{west}
D.~West \emph{et~al.}, ``Water-scrum-fall is the reality of agile for most
  organizations today,'' \emph{Forrester Research, July}, vol.~26, 2011.

\bibitem{doingtobeing}
P.~Ranganath, ``Elevating teams from {Doing} agile to {Being} and {Living}
  agile,'' in \emph{Agile Conference (AGILE), 2011}, Aug 2011, pp. 187--194.

\bibitem{williams}
L.~Williams, ``What agile teams think of agile principles,''
  \emph{Communications of the ACM}, vol.~55, no.~4, pp. 71--76, 2012.

\end{thebibliography}

\end{document}